\documentstyle[prl,aps,multicol,epsf]{revtex}
\begin{document}
\draft
\widetext
\title{Electron teleportation with quantum dot arrays}
\author{Olivier Sauret$^a$, Denis Feinberg$^{a}$, Thierry
Martin$^{b}$}
\address{$^a$Laboratoire d'Etudes des Propri\'etes
Electroniques des Solides, CNRS, BP166, 38042 Grenoble, France}
\address{$^b$
Centre de Physique Th\'eorique et  Universit\'e de la
M\'editerran\'ee, Case 907, 13288 Marseille, France}
\maketitle
\widetext
\begin{abstract}
An electron teleportation protocol, inspired by the scenario 
by Bennett et al., is proposed in a mesoscopic set-up.  
A superconducting circuit allows to both 
inject and measure entangled singlet electron pairs in an array of three
normal quantum dots.  The selection of the teleportation process is
achieved in the steady state with the help of two superconducting dots and
appropriate gating.  Teleportation of the electron spin is detected by
measuring the spin-polarized current through the normal dot array.  This
current is perfectly correlated to the pair current flowing inside the
superconducting circuit. The classical channel required by Bennett's
protocol, which signals the completion of a teleportation cycle, 
is identified with the detection of an electron charge in 
the superconducting circuit. 
\end{abstract}
\begin{multicols}{2}
\narrowtext


Teleportation (TP) recently entered the realm of quantum 
physics when Bennett et al. \cite{bennett} proposed a protocol to reconstruct
the unknown state of a given
particle at a different location.  The sender, Alice, and the receiver,
Bob, share an entangled pair \cite{EPR} -- , and
Alice performs a joint measurement on the ``source'' particle and
her part of the pair.  The result of the measurement is
communicated through a classical channel to Bob, allowing him to
reconstruct the initial state on his part of the pair.  This
protocol has since been experimentally demonstrated with
polarized photons \cite{bouw}, as well as proposed in atomic
physics \cite{bose} and solid state optics \cite{excitons}. 
Besides its fundamental character, TP is likely to become an
essential element of future information processing schemes
\cite{q_information}.  It is certainly relevant to test these
manifestations of non-locality \cite{photons} with massive particles in
nanostructured devices, with the advantage that these can be integrated in
(quantum) electronic circuitry. Similar analogies
between photon propagation and phase-coherent electron
transport in nanostructures were illustrated by the fermion version
of the Hanbury-Brown and Twiss experiment \cite{HBT}.

The general principle of the present mesoscopic scheme for TP
-- an array of quantum dots with superconductors -- is inspired
of Ref \cite{bennett}, but follows more closely its optical
implementation \cite{bouw}.  Alice's
measuring device for entangled (singlet) electron pairs is an s--wave
superconductor, as is the generator of the entangled electron pairs
\cite{choi_bruder_loss,loss,lesovik_martin_blatter,feinberg}. 
Similarly to the  the optics
experiment only one of the four Bell states is measured.

However photons interact weakly (except during 
their generation and detection process).  On the contrary, electrons in
nanostructures experience strong Coulomb interactions, which can be used to
ensure that electrons be injected one by one from/to a quantum dot
through tunnel barriers
\cite{devoret}. Indeed, further control can be obtained in a
multidot array, by means of intradot and interdot Coulomb correlations:
here, the ``correct'' TP sequence (injection, pair creation, measurement,
classical channel and detection) can be precisely selected, while operating
in the steady state, by an appropriate initial choice of gate voltages. 

\begin{figure} \centerline{\epsfxsize=7cm 
\epsfbox{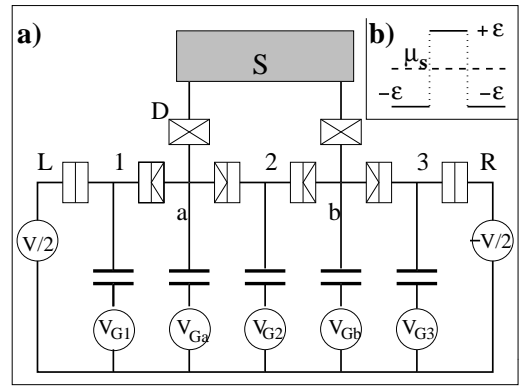}}
\vspace{4mm}
\caption{a) The TP cell contains: i)
NN junctions between reservoirs $\bf L,\bf R$ and dots $\bf 1$ and $\bf 3
$; ii)
N-S junctions between $({\bf 1},{\bf a})$, $({\bf a},{\bf 2})$, $({\bf
2},{\bf b})$ and $({\bf b},{\bf 3})$, and S-S junctions between $\bf a$
($\bf b$) and the bulk superconductor $\cal{S}$.  Capacitive couplings
select the correct sequence. Detector $D$ (i.e. capacitive coupling betweeen 
$\bf a$ and $\cal S$) signals the passage of a Cooper pair 
in the superconducting circuit. b) Sketch of the energy 
level configuration of
dots $\bf 1$, $\bf 2$ and $\bf 3$ (electron energies $\pm
\varepsilon$ are symmetric with respect to the
superconductor chemical potential $\mu_S$).
\label{fig1}} \end{figure}

The ``device'' is depicted in Fig.  \ref{fig1}a: three normal (N) dots,
$\bf 1$, $\bf 2$ and $\bf 3$, and two superconducting (S) dots
$\bf a$, $\bf b$, are placed in alternation: N-dots can
only communicate via the S-dots. Dots $\bf 1$ and $\bf 2 $ are coupled to
dot $\bf a$ -- Alice's measuring device -- by tunnel junctions, while 
$\bf 2$ and $\bf 3$ are coupled to $\bf b$ -- the source of entangled
pairs.  Furthermore, dots ${\bf a}, {\bf b}$ are connected by tunnel junctions to a
superconducting ($\cal{S}$) circuit where Cooper pairs only are transferred.  
Reservoir $\bf L$ emits in dot $\bf 1$ the electron to be teleported, 
and reservoir $\bf R$ (``Bob'') collects
the teleported state from dot $\bf 3$.
First, focus on the isolated system of $5$ dots.
Following \cite{bennett},
an entangled singlet pair of particles 
$|\Psi^S\rangle_{23} = 2^{-1/2}(|\uparrow
\downarrow\rangle_{23} - |\downarrow
\uparrow\rangle_{23})$ is produced by
$\bf b$.  Coulomb blockade \cite{devoret} prohibits
double occupancy in each dot
\cite{choi_bruder_loss,loss} (the same is true for Cooper pair
occupancy in the superconducting dots). Bringing together the singlet
$|\Psi^S\rangle_{23}$ with the state
$|\sigma\rangle_1$ to be teleported, the resulting state with dots $\bf
1,\bf 2,\bf 3$ occupied leaves the spin in $\bf 3$ unspecified.  This
three-particle wave function is now decomposed among the $4$ Bell states
for electron spins in dots ($\bf 1$, $\bf 2$)
\cite{bennett}: $ |\Psi\rangle_{123}=
-(1/2)[|\Psi^S\rangle_{12}|\sigma\rangle_3 +
\sum_{\nu}|\Psi^{T_{\nu}}\rangle_{12}
|\tilde\sigma_{\nu}\rangle_3]$
where
$|\tilde\sigma_{\nu}\rangle$ are unitary transforms of
$|\sigma\rangle$ ($\nu =0,\pm$) and
$|\Psi^{T_{0+-}}\rangle_{12}$ the three triplet states.
$\bf a$ acts as a detector for the singlet state of
electrons in $\bf 1$ and $\bf 2$: absorption of a
Cooper pair only occurs if $(\bf 1,\bf 2)$ contain a singlet. 
The remaining spin in dot
$\bf 3$ necessarily acquires the same state $|\sigma\rangle$ as
the initial spin in dot $\bf 1$, as required by TP. The
absorption of the singlet electron pair from ($\bf 1$, $\bf 2$)
also destroys the initial spin state in dot $\bf 1$, therefore
satisfying the ``non-cloning theorem'' \cite{zurek}, and this transition
is made irreversible because it is followed
by the  (irreversible) injection of a ``new'' electron from reservoir $L$.

A microscopic model supports this TP protocol.  N-dots are assumed to have a
discrete spectrum, with level spacing comparable to the gaps
$\Delta_{a,b,\cal{S}} \sim
\Delta$ of the S-dots and S-circuit.
The S-dots have a continuous quasiparticle spectrum,
and $\Delta_{a,b}>E_{C_\mu}\equiv e^2/C_{\Sigma_\mu}$.
Only two occupation numbers are kept for each dot.
N-dots (S-dots) have ``empty''
states with an even  number $N^{0}_{\mu}$ of electrons,
and have ``filled'' states  with $N^{0}_{\mu}
+1$  ($N^{0}_{\mu}+2$) electrons.
The Hamiltonian which describes the
TP cell reads $H=H_0 + H_t + H_C$ where $H_0$
describes the isolated elements (dots and reservoirs). 
The single electron hopping term $H_t$ has
amplitudes $t_{\alpha\beta}$ ($\alpha,
\beta=\{\bf L,\bf R,\bf S,\bf 1,\bf 2,\bf 3,\bf a,\bf b\}$). 
Only one level is relevant in each N-dot, and next nearest
neighbor hoppings are neglected.  The Coulomb contribution has the standard
form: $ H_C=(1/2)\sum_{\mu,\nu=1,2,3,a,b}U_{\mu \nu}
\delta N_{\mu} \delta N_{\nu}$,
where
$\delta N_{\mu}=N_{\mu}-\bar{N}_{\mu}$
is the deviation from the effective number of electrons
imposed by the gates (voltage $V_{G\mu}$). The coefficients $U_{\mu \nu}$
form the inverse capacitance matrix of this five dot system, and are
computed \cite{inpreparation} from the individual capacitances
$C$,$C'$, $C_s$ and $C_g$ of the NN, NS, SS and gate junctions
respectively (see Fig.  1a).  The dots are coupled to the N/S
reservoirs with energy line widths $\Gamma_{L,R} = 2\pi
t_{L1(R2)}^2 N_{L(R)}(0)<<\Delta$, with density of states $N_{L,R}(0)$
(and similarly $\Gamma_{Sa} =\Gamma_{Sb}$). The chemical
potential $\mu_S$ of the superconductor is located in the
middle of the left/right reservoir potentials $\mu_S\pm eV/2$.
Dot configurations are identified by the occupation numbers
of dots $\bf 1,\bf a,\bf 2,\bf b,\bf 3$: $0$ or $1$ ($0$, $1$
or $2$) for the N-dots
(S-dots). Charging energy differences $\Delta
E_{i}^{f}$ between the initial and final configurations of the
five dot system enter the $O(H_t)^2$ calculation of the pair
tunneling amplitude from $b$ to
$2,3$ (and similarly from $\bf a$ to($\bf 1$, $\bf 2$)): $A_P^{b}\simeq
2\sum_{k,x} u^{b}_k v^{b}_k t_{2b}t_{3b}/ (i\eta-E_k^{b}-\Delta
E_{00020}^{00x0\bar{x}}) $, with $\eta$ an infinitesimal, $u_{k},v_{k}$ the
usual BCS parameters, $x=0,1 (\bar{x}=1,0)$.  $E_{k}$ is the quasiparticle
energy involved in the creation of a quasiparticle.  The amplitude
$A_P^{a(b)}$ is at most comparable to $\Gamma_{Sa,b}$, and decreases with
the distance between the two junctions involved in cross Andreev reflection
\cite{choi_bruder_loss,falci}.  The transition amplitude $A_S$ between $\bf
a$($\bf b$) and the $\cal{S}$-circuit is in general larger \cite{matveev}. 
Consider the system in the absence of connections with the N,S leads.  Dot
gate voltages are adjusted so that the  pair transitions $A_P^{a,b},
A_S^{a,b}$ are resonant.  Discarding virtual processes with more than one
quasiparticle in $\bf a$ or $\bf b$, one obtains the effective pair
Hamiltonian
\begin{equation}
H_{eff} = A_P^a \Psi^{\dagger}_{12} \Psi_a + A_P^b
\Psi^{\dagger}_{23}\Psi_b + A_S (\Psi^{\dagger}_a +
\Psi^{\dagger}_b)\Psi_{\cal S} +H.c.
\label{effective Hamiltonian}
\end{equation}
where the $\Psi_{\alpha\beta}$ destroys a singlet pair in $2$
N-dots and $\Psi_{a,b,{\cal S}}$ destroys a Cooper pair in the
superconducting elements.

\begin{figure}
\centerline{\epsfxsize=8cm
\epsfbox{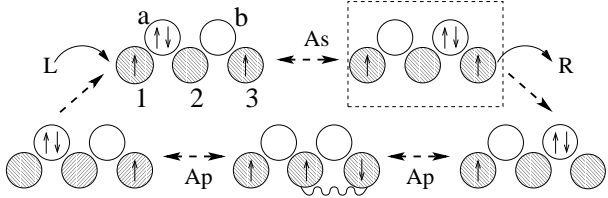}}
\vspace{4mm}
\caption{
TP sequence: upper dots (white)
are  S-dots $\bf a$ and $\bf b$, lower dots (shaded) are N-dots
$\bf 1$, $\bf 2$, $\bf 3$. Horizontal transitions only are
resonant. Starting from the framed configuration (upper right), an electron 
in $\bf 3$ escapes in $\bf R$; next, a pair (from $\bf b$) creates 
an entangled state ${\bf 2},{\bf 3}$ (wiggly line) with rate $A_P$, 
leaving all N-dots filled.  A pair ${\bf 1},{\bf 2}$ then escapes in ${\bf a}$. 
The electron in ${\bf 3}$ acquires the spin state of dot ${\bf 1}$, as confirmed
by the absorption of a signet state in ${\bf a}$ and the subsequent
injection of an electron from ${\bf L}$.
\label{fig2}}  \end{figure}

The TP sequence is now illustrated (Fig. \ref{fig2})
in a steady state operation of the whole circuit (``TP cell''),
by applying a constant bias between reservoirs $\bf L$ and $\bf
R$. Circuit parameters and gate voltages are chosen such that
the TP cell is symmetric in changing $\bf 1$ ($\bf a$) into $\bf 3$ ($\bf
b$), thus $A_P^a = A_P^b$ (no phase difference exists in the S part of the
cell).  The TP sequence repeats
itself cycle after cycle, each one achieving teleportation of an electron
injected in $\bf 1$ from $\bf L$, and detection in $\bf R$ of the teleported
electron in $\bf 3$.  Start with dots $\bf 1$, $\bf 3$ and $\bf b$ occupied
(upper right in Fig. \ref{fig2}).
The teleportation process is triggered by the escape of the electron in $\bf 3$
in reservoir $\bf R$. 
Doing so, the energy level in $\bf b$ is lowered, 
thus interrupting the previously resonant Cooper pair transfer.  
Now (lower right) the pair in $\bf b$ resonates with ($\bf 2,\bf
3$), building with $\bf 1$ the aforementioned state $|\Psi\rangle_{123}$.
Measurement of the singlet state in ($\bf 1,\bf 2$) by $\bf a$ (Alice) is
achieved when a new electron is injected into $\bf 1$, 
yet also raising the energy level in $\bf a$
in the process.  The remaining electron in $\bf 3$ thus
irreversibly acquires the state of the previous one in $\bf 1$, while the
new electron waits in $\bf 1$ to be teleported in the next sequence.

Note that : i) Incoherent processes are brought by the
reservoirs and the applied bias. The latter bias also 
determines the direction of TP (right or left) in an
otherwise symmetric TP cell.  This allows pair production from $\bf b$ and
pair measurement in $\bf a$ to be both irreversible. ii) Successive TP cycles
are linked together in such a way that a detection event triggers pair
production for the next cycle, and an injection event triggers pair
measurement for the previous cycle. iii) The classical channel corresponds 
to the detection of  an extra Cooper pair in the superconducting circuit
(${\bf a}+{\cal S}+{\bf b}$). 
In Fig. \ref{fig1}, this detection is schematized 
by the presence of a detector $D$, positioned between $\bf a$ and 
$\cal S$. 
It conveys the information about the (classical) charge in dot $\bf a$,
this signals the completion of the pair transition $1,2 \to a$, the destruction of 
the original (quantum) spin state in dot $1$ and the
instantaneous reconstruction of this state in dot $3$.  This is enabled by
the entangled pair, in full agreement with the TP principle \cite{bennett}. 
iv) This classical information should in principle be transmitted to Bob,
in order to distinguish whether the electron he receives from $\bf 3$ is
the result of a TP or any other transport process.  Here, this would require a
time resolved correlation measurement between the current in the S-circuit
and that injected in $\bf R$, i.e. the analog of coincidence measurements
performed in the optical implementation \cite{bouw}.  Yet, the present
mechanism has the merit of automatically implementing TP in the five-dot
cell; v) As in optics
\cite{bouw,bouw2}, measurement of the sole singlet state reduces to
$1/4$ the efficiency of TP, but not the fidelity, equal to $1$
in the ideal sequence depicted above.

The sequence reads: $...10021] \rightarrow
[ 10020 \leftrightarrow 10101 \leftrightarrow 02001 ] 
\rightarrow [ 12001 \leftrightarrow 10001
\leftrightarrow 10021 ]...$
Close inspection of the energy balance $\Delta E_{i}^{f}$ of all the electronic
transitions in the TP cell reveals that it is indeed possible to force this
sequence with the help of constant gate voltages only \cite{inpreparation}. 
As an example, let us assume that $C=C'=C_{s}=100 C_{g}$.  First, the
resonance condition for pair transitions implies $\Delta
E_{10020}^{10101}=\Delta E_{10101}^{02001}=\Delta E_{12001}^{10001}=\Delta
E_{10001}^{10021}=0$.  One finds that it fixes $\bar{N}_{a,b}= 0.97$,
 and $\bar{N}_{1} +
\bar{N}_{2}=0.67$.  Second, injection and detection are ensured (with
$\mu_{L,R} =\pm eV/2$) by $\Delta E_{02001}^{12001} < eV/2$, $\Delta
E_{10021}^{10020} < eV/2$, therefore $V >(\bar{N}_{1}-0.9)e/C $.  Third,
the transfer of an electron from $\bf 3$ to $\bf R$ is allowed from state
$10021$ but, among other unwanted transitions, not from $10101$ or $02001$: 
this can be achieved in a certain range of $V$ because $\Delta
E_{10101}^{10100} -
\Delta E_{10021}^{10020} =2U_{b3} - U_{23} =11e^2/30C > 0$ and $\Delta
E_{02001}^{02000} -
\Delta E_{10021}^{10020} = U_{13} + 2U_{b3}-2U_{a3} = 13e^2/30C > 0$.

TP fidelity is reduced by other transport processes, yet which are 
suppressed by our choice of resonant Cooper pair transfers.  First, a direct
electron transfer can result from two consecutive cotunneling transitions
from dot $\bf 1$ to dot $\bf 2$, and from dot $\bf 2$ to $\bf 3$, while
generating virtual quasiparticles
\cite{falci,cotunneling}.  Cotunneling can be avoided by
maximizing the energy differences for transitions from dot $\bf 1$ to dot
$\bf 2$, by tuning the parameter $\bar{N}_{1} -
\bar{N}_{2}$.  Positive (negative) gate voltages applied to dots $\bf 1$,
$\bf 3$ (dot$\bf 2$) guarantee that cotunneling involves a positive energy
$2\varepsilon$ (Fig.\ref{fig1}), with $A_P \ll
\varepsilon < \Delta$.  The amplitude for cotunneling from dot
$\bf 1$ to $\bf 3$ is reduced as it scales like
$A_P^2/\varepsilon \ll A_P$. Cotunneling is quenched by
maximizing $\Delta E_{12001}^{02101}$, $\Delta
E_{10021}^{00121}$, $\Delta E_{10001}^{00101}=
(\bar{N}_{1}-8/15)e^2/C\sim 2\varepsilon$.  At $T =0$, optimal
operation is obtained with $\bar{N}_{1} = \bar{N}_{a} \sim 1$,
$\bar{N}_{2} \sim -1/3$ and finite bias $0 < V < e/3C$. A second
process is Josephson tunneling between $\bf a$ and $\bf b$,
independently of the pair current involved in the TP sequence: 
Cooper pairs can be transmitted by cotunneling through
dot $2$ only. However, this process
\cite{choi_bruder_loss,glazman} involves 
quasiparticle excitations in $\bf
a$ or/and $\bf b$, contrary to the TP process. Note that TP
involves a (normal) spin-conserving current between $\bf L$ and
$\bf R$, perfectly correlated to a pair current in the $\cal{S}$
circuit. This signature of the coupled quantum and classical
channels allows to distinguish TP from the other two processes :
the pair current is missing in cotunneling, while the normal
current is missing in the Josephson process.

Each TP cycle may involve a new spin state, independent of the previous
one.  Yet it is convenient, to test TP, to fully spin polarize both $\bf 
L$ and $\bf R$, in order to measure the spin correlation
between the incoming
and outgoing electrons (similarly to the optics experiment
\cite{bouw}).  Assuming the TP cell to be weakly coupled to the reservoirs,
transport across the dot array can be described by a master equation. 
Defining states $|\uparrow,2,0,0,\uparrow\rangle =|a\rangle$,
$|\uparrow,0,0,0,\uparrow\rangle =|c\rangle$,
$|\uparrow,0,0,2,\uparrow\rangle =|b\rangle$, $|\uparrow,0,0,2,0\rangle
=|1\rangle$, $|0,2,0,0,\uparrow\rangle =|3\rangle$, and $|S\rangle$,
$|T\rangle$ the states $|10101\rangle$ with wave functions
$|\Psi^S\rangle_{12}|\sigma\rangle_3$ and
$(1/\sqrt{3})\sum_{\nu}|\Psi^{T_{\nu}}\rangle_{12}
|\tilde\sigma_{\nu}\rangle_3$, the Bloch equations  
for the reduced density matrix, 
describing both the
populations and the coherences $\sigma_{\mu \nu}$ ($\mu, \nu = a,b,c,1,2,3$)
can be written in the general form
\cite{sequential,gurvitz} at zero temperature:
\begin{eqnarray}
\dot\sigma_{\mu \mu} &=&
i\sum_{\nu}\Omega_{\mu \nu}(\sigma_{\mu \nu}-\sigma_{\nu \mu})
- \sum_{\lambda}(\Gamma_{\mu \lambda}\sigma_{\mu \mu}
- \Gamma_{\lambda\mu}\sigma_{\lambda \lambda})\\
\dot\sigma_{\mu \nu} &=& i\sum_{\lambda}(\sigma_{\mu\lambda}
\Omega_{\nu\lambda} -\sigma_{\lambda
\nu}\Omega_{\mu \lambda}) -\frac{\sigma_{\mu \nu}}{2}
\sum_{\lambda} (\Gamma_{\mu \lambda} +  \Gamma_{\nu \lambda})
\end{eqnarray}
with $\Omega_{ac} =\Omega_{ca} =\Omega_{bc} =\Omega_{cb} =
A_S$, the tunneling rate for Cooper pairs from $\bf a$ to $S$ ($S$
to $\bf b$). $\Omega_{1S} =\Omega_{S1} =-A_P/2$,
$\Omega_{1T} =\Omega_{T1} =-\sqrt{3}A_P/2$,
$\Omega_{3S} =\Omega_{S3} =A_P$, $\Gamma_{b1} =\Gamma_R$,
$\Gamma_{3a} =\Gamma_L$, all the other $\Omega_{\mu \nu}$'s and
$\Gamma_{\mu \nu}$'s are zero. 
The steady state TP
current $I_{tel} =e \Gamma_R \sigma_{bb}$ (from $L$ to $R$)
then reads:
\begin{equation}
I_{tel} =e \frac{\Gamma_L
\Gamma_R}{(\Gamma'_L+4\Gamma_R)}
\frac{A_P^2}{[A_P^2 + 2\Gamma_L^2
\Gamma_R/(\Gamma'_L+4\Gamma_R)]}
\end{equation}
with $\Gamma'_L =\Gamma_L (3 + \Gamma_R^2/2A_S^2)$.
Note that the above analysis does not depend on the 
incident polarization as depicted in Fig. \ref{fig2}, as the two
spin channels are totally decoupled. 
Aside from corrections due
to cotunneling, the only transport channel through
the dot array is the TP process.  Unless direct evidence comes
from individual electron coincidence counting (as for photons), a signature
of TP is already provided by the equality of the TP current and the pair
current $I_P=2I_{tel}$.

As in quantum optics, a
proof for nonclassical spin correlations requires to check the above equality
for parallel spin polarizations of reservoirs $\bf L$ and $\bf R$, taking
successively two values corresponding to non-orthogonal quantum states
\cite{bennett}.  Refined diagnosis for TP can be searched in noise
correlations measurements\cite{HBT,lesovik_martin_blatter} or with future
time-resolved measurements.

Limiting factors are now
considered.  First, it is crucial to maintain spin coherence during the
TP sequence (on a time scale $\sim\hbar/\Gamma_{R,L}$, which turns out to 
be ``short'' in practice).  
This coherence can be destroyed by spin-orbit coupling, 
or by collisions with the other electrons within the
dot.  Such spin-flip processes can be minimized provided that the level
spacing in the dots is larger than the temperature and the resonance width
of the dots\cite{loss}: ``empty'' dot states of $\bf 1,\bf 2,\bf 3$ 
should preferably
have even filling $N_{\mu}$.  Second, the present scheme is clearly
optimized if Cooper pair transfer from the N-dots pairs to each S-dots has
an maximal amplitude $A_P$.  This amplitude is strongly reduced by a
geometrical factor in two and three dimensions when the two N-S tunnel
barriers as spaced farther than a few nanometers
\cite{choi_bruder_loss,loss,falci}.
On the other hand, the size of the
S-dots is large enough so that $E_{C_{a(b)}} < \Delta$, thus precise
lithography bringing N-dot pairs close together (however avoiding direct
tunneling between N-dots) is required.  An alternative would be to define
the dots with quasi one--dimensional conductors (nanotubes) placed in
contact with the superconductor, as the geometrical constraint is relaxed
\cite{recher_loss}.

To sum up, an electron spin teleportation scheme which employs
N-S hybrid nanodevices for electrons is proposed.
It relies on current nanofabrication techniques and operates in
the steady state, using Coulomb correlations in the dot array.  The TP 
current which flows between the 
N-reservoirs is locked with the
pair current in the
S-circuit.
The device could be implemented using bent, gated, contacted
carbon nanotubes next to superconductors
\cite{bezryadin}.  The feasibility of
this proposal relies on efficient spin filters $ L$, $R$, already
available at low temperatures \cite{spinfilter}.  Finally, the
proposed setup is generalizable to $2{\cal N} +1$ normal dots,
together with ${\cal N}$ superconducting circuits ($2{\cal N}$
S-dots): TP of a spin state in  dot $\bf 1$ onto dot $2{\cal N}
+1$ can be achieved by a swapping process
\cite{q_information}, thus extending the range of TP.

Stimulating discussions with V. Bouchiat are
gratefully acknowledged.
LEPES is under contract with Grenoble universities, UJF and INPG.

\end{multicols}
\end{document}